\documentclass{article}
\usepackage[T1]{fontenc} % add special characters (e.g., umlaute)
\usepackage[utf8]{inputenc} % set utf-8 as default input encoding
\usepackage{amsmath,cite,url}
\usepackage{graphicx}
\usepackage[lbd]{ismir}
\usepackage{gensymb}

\usepackage{color}

\usepackage[firstpage]{draftwatermark}
\definecolor{lightgray}{rgb}{0.9,0.9,0.9}
\definecolor{darkgray}{rgb}{0.4,0.4,0.4}
\SetWatermarkFontSize{12pt}
\SetWatermarkScale{1.1}
\SetWatermarkAngle{90}
\SetWatermarkHorCenter{202mm}
\SetWatermarkVerCenter{170mm}
\SetWatermarkColor{darkgray}
\SetWatermarkText{Late-Breaking / Demo Session Extended Abstract, ISMIR 2025 Conference}

\DeclareMathOperator*{\argmax}{arg\,max}

\title{Live Vocal Extraction from K-pop Performances}

\threeauthors
  {Yujin Kim} {New York University \\ {\tt yk3281@nyu.edu}}
  {Richa Namballa} {New York University \\ {\tt rn2214@nyu.edu}}
  {Magdalena Fuentes} {New York University \\ {\tt mf3734@nyu.edu}}

\def\authorname{Y. Kim, R. Namballa, and M. Fuentes}

\begin{document}

\maketitle
\begin{abstract}
K-pop’s global success is fueled by its dynamic performances and vibrant fan engagement. 
Inspired by K-pop fan culture, we propose a methodology for automatically extracting live vocals from performances.
We use a combination of source separation, cross-correlation, and amplitude scaling to automatically remove pre-recorded vocals and instrumentals from a live performance. Our preliminary work introduces the task of live vocal separation and provides a foundation for future research in this topic. 
\end{abstract}

\section{Background}\label{sec:background}

It is common in popular music for vocalists to perform live with a pre-recorded backing track that could contain anything from instrumentals to additional main and back-up vocals, all sent to their in-ear monitor to help them match pitch and harmonize correctly.
From the listener's perspective, the backing track makes the live vocals sound more full and pleasing to the average ear, especially in K-pop.

K-pop is a form of popular music distributed by entertainment agencies in South Korea.
These companies specialize in training and managing artists (\emph{idols}) to execute rigorous choreography during high-energy performances which blend elements of various well-known genres (e.g., hip-hop, dance music) together in unique ways~\cite{Lee2024}.
This variety has allowed K-pop to skyrocket in global popularity during the second ``Korean wave'' (\emph{hallyu} 2.0), empowered by digital streaming and social media engagement~\cite{Lee2015}.

Fans often create videos of idol performances with the music recording and backing track removed (``MR Removed'') to share them on platforms, such as YouTube, as a way of showing their support for their favorite groups.
Fans want to hear the artists' live singing, so they will ``extract'' the live vocals from the playback track using a digital audio workstation (DAW).
A key factor in the feasibility of this process is that the idols seldom sing with a live accompaniment on Korean music shows, so the broadcasted versions sound similar to the released album track and are typically matched in tempo.
We explore how to automate this manual process of live singing vocal extraction using simple signal processing and pretrained source separation models.
There is significant discourse in the K-pop community about the merit of these types of ``MR Removed'' videos as they are often weaponized in fandom wars to accuse artists of lip-syncing.
We look at this approach solely to understand its effectiveness as a methodology for live vocal extraction, with no intention of inciting a controversy in the K-pop industry.

\section{Related Work}\label{sec:related_work}

There has been remarkable improvement in the task of singing voice separation over the years with the use of deep learning techniques~\cite{Casey2000, Jang2003, Vembu2005, Virtanen2006, Smaragdis2014,  Jansson2017, Rouard2023, Luo2023, Lu2024}.
While the results of these models are impressive, they usually group all singing voices together in the outputted stem, making no differentiation between the main and backing vocals.
Furthermore, when processing recordings of live performances, the models do not distinguish between portions that are pre-recorded and the parts sung live.
Understandably, this novel problem appears quite difficult to address given the granularity of the separation required.
Therefore, we hope that our work sets the foundation for future research investigating the challenge.

Additionally, despite the historical dominance of Western music in MIR, several prior studies have leveraged K-pop in their research\cite{Hu2014, Kim2017, Lee2020, Kim2024, Park2024}.
K-pop is notable for its song structure being driven by the performance (choreography) and group members' roles~\cite{Lee2024}.
It is industrial in that the songs are systematically produced, but customizable to each group's aesthetics.
This precise nature of K-pop motivates our methodology.

\begin{figure*}
    \centering
    \includegraphics[alt={A flow diagram depicting the live vocal extraction pipeline for K-pop performances.
    The audio from the live performance and a copy of the released song are passed through the Demucs source separation model to separate each version into its vocals and accompaniment.
    The optimal lag is computed using the instrumentals to align the vocal signals, which are them matched in amplitude through scaling.
    Minor lag adjustments are made and then the recorded vocals are subtracted from the live performance resulting in the extracted live vocals as the residual waveform.}, width=0.95\linewidth]{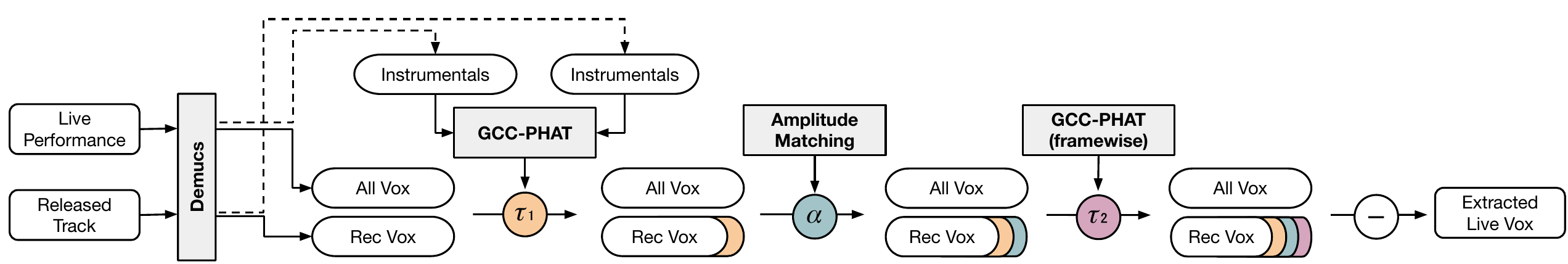}
    \caption{Pipeline overview for the live vocal extraction process.
    The pre-recorded and live performance versions of a K-pop song are separated into vocal and accompaniment stems.
    The stems are aligned and matched in amplitude before subtracting the recorded vocals from the live performance vocal signal, resulting in the live vocals as the residual waveform.}
    \label{fig:process}
\end{figure*}

\section{Methodology}\label{sec:methodology}

\subsection{Manual Baseline}\label{subsec:baseline}

For the baseline method, we replicated the original, simplest approach and extracted the live vocals in a DAW (Avid Pro Tools) by following a fan-made tutorial\footnote{\scriptsize\url{https://www.youtube.com/watch?v=WJWMCJBZssM}}.
We tried this method first to compare the separation quality between DAW-based workflows and automated signal processing techniques.
First, we imported a recording of the live performance and a copy of the released track.
All audio was stored in WAV format at 44.1kHz/16bit.
We then manually aligned the two tracks by looking at samples and matched the amplitudes by adjusting the gain through trial-and-error.
The pre-recorded audio waveform was inverted $180\degree$ and mixed with the live performance signal to cancel out the overlapping components.
Ideally, only the live vocal components from the performance would remain.

\subsection{Proposed Implementation}\label{subsec:implementation}

For our proposed implementation, we wanted to both automate and improve the live vocal extraction.
First, both the live performance audio and released track were passed through HT Demucs~\cite{Rouard2023} to separate the vocal and accompaniment stems and then converted to mono.

To align the two signals, we leveraged the hypothesis that the accompaniments for both versions of the song should be nearly the same. 
We computed the Generalized Cross-Correlation with Phase Transform (GCC-PHAT) with the Discrete Fourier Transform ($\mathcal{F}$), to estimate the time delay between the two real-valued instrumental signals $\mathbf{x}$ and $\mathbf{y}$, allowing a maximum shift of $\pm$20s.
Then, we found the optimal lag $\tau^{*}$ (in samples) which maximized the correlation and applied it to shift the vocal stem of the released track to match the onset of the live performance vocal stem.
The stems were zero-padded to match in length.

\vspace{-0.5cm}
\begin{equation}
    \tau^{*} = \argmax_{\tau}\, \left( \mathcal{F}^{-1}\left( \frac{\mathcal{F}(\mathbf{x}) \cdot \overline{\mathcal{F}(\mathbf{y})}}{| \mathcal{F}(\mathbf{x}) \cdot \overline{\mathcal{F}(\mathbf{y})} |}\right) [\tau] \right)
\end{equation}

To accurately cancel out the pre-recorded signal from the live performance signal, we computed the scale factor ($\alpha$) to adjust the gain of the pre-recorded vocal signal to match the live one.
We calculated the framewise Pearson correlation between the two signals using a frame length of 1.0s, hop size of 0.5s, and Hann window.
The frame with the highest correlation is then utilized to perform a linear least-squares optimization to find the optimal $\alpha^{*}$ to match amplitudes between the stems in that frame:
\begin{equation}
    \alpha^{*} = \min_{\alpha} \| \alpha \cdot \mathbf{y_{\text{rec}}} - \mathbf{y_{\text{live}}} \|^2 = \frac{\mathbf{y_{\text{rec}}}^\mathsf{T} \mathbf{y_{\text{live}}}}{\|\mathbf{y_{\text{rec}}}\|^2}
\end{equation} 
The value $\alpha^{*}$ is applied to the full pre-recorded vocal stem.

Lastly, we compute the GCC-PHAT between the amplitude-matched vocal stems to see if any minor lag adjustments need to be made.
We do this framewise, but limit the maximum shift to 0.25s, take the most frequent lag value across all frames, and apply it to the entire signal.
Finally, we subtract the adjusted pre-recorded vocal stem from the live performance vocal stem to get the residual waveform, which we call the extracted live vocals.
The overall process (Figure~\ref{fig:process}) is automated for user convenience and the implementation is publicly available\footnote{\scriptsize\url{https://github.com/yujin-kimmm/live-vox-extraction}}.

\section{Results}\label{sec:results}

We applied both the manual baseline approach and our proposed implementation on songs from five well-known idol groups using their performances on popular Korean music shows: ``Lovesick Girls'' (BLACKPINK), ``I Can't Stop Me'' (TWICE), ``REALLY REALLY'' (WINNER), ``BANG BANG BANG'' (BIGBANG), and ``HARD'' (SHINee).
Since we do not have the ground-truth live vocal stems for these songs, our evaluation was done qualitatively through listening tests by the authors.

Our implementation is fast and requires little manual work compared to the DAW approach.
We found our method performed best when the live performance and the released song properly aligned during our initial lag calculation. The pre-recorded backing vocals become noticeably attenuated, making the live vocals clearer, such as in the BLACKPINK and TWICE songs.
Using Demucs to separate the vocal stems first has a major advantage over using only the DAW, as it minimizes instrumental interference in the alignment and scaling process.

Despite aligning the audio, sometimes the pre-recorded vocals would not cancel and made the live vocals sound fuller with a chorus effect.
This misalignment could be attributed to crowd noise and vocal reverberation, as shown in ``REALLY REALLY'' and ``HARD''.
In this case, the resulting audio sounds similar to the Demucs output.

Our results also showed even the slightest tempo mismatch would nullify the effectiveness of our approach. This occurred in ``BANG BANG BANG'' where the live performance was slower than the album version by only a few beats. Properly aligning the waveforms would require the use of time-stretching, which is not currently included in our implementation.

\section{Conclusion}
We introduced the task of live vocal extraction in the context of K-pop performances.
Our automated approach uses deep learning and signal processing techniques, but we hope to further expand the study by implementing more intelligent separation models.

\bibliography{references}

\end{document}